\documentclass[11pt,a4paper,copyright,goog]{google}

\usepackage[numbers,compress]{natbib}
\usepackage{xurl}
\usepackage{hyperref}
\usepackage{multirow}
\usepackage{bm}
\usepackage{quoting}
\usepackage{ragged2e}
\usepackage{array}
\usepackage{enumitem}
\usepackage{rotating}
\usepackage{moresize}
\usepackage{float}

\usepackage{booktabs}      
\usepackage{array}         
\usepackage{caption}       
\usepackage{graphicx}
\usepackage{subcaption}
\usepackage{chngcntr}
\usepackage{longtable}
\usepackage{listings}
\usepackage{xcolor}
\usepackage{xspace}
\usepackage{cleveref}

\definecolor{backcolour}{rgb}{0.95,0.95,0.92}
\lstdefinestyle{monospacedgrey}{
    backgroundcolor=\color{backcolour} \& 
    basicstyle=\ttfamily\scriptsize \& 
    numbers=none \& 
    breaklines=true \& 
    breakatwhitespace=true \& 
    postbreak=\mbox{\textcolor{red}{$\hookrightarrow$}\space} \& 
}
\crefformat{footnote}{#2\footnotemark[#1]#3}

\newcolumntype{C}[1]{>{\centering\let\newline\\\arraybackslash\hspace{0pt}}m{#1}}

\newcommand{\takeout}[1]{}

\newcommand{\figref}[1]{Figure~\ref{#1}}

\newcommand{\agbib}[1]{}
\newcommand{\secref}[1]{Section \ref{#1}}
\newcommand{\commentout}[1]{}

\newcommand{\amirgc}[1]{\textcolor{brown}{AG: Cathy PTAL.{#1}}}

\newcommand{\ignore}[1]{}
\usepackage{xcolor}

\paperurl{}

\uselogo{} 

\title{Improving Entrepreneurial Skills by Practicing with AI}
\author[a]{Amir Globerson}
\author[a]{Anisha Choudhury}
\author[a]{Anna Iurchenko}
\author[b]{Ashley Hemmy}
\author[a]{Avinatan Hassidim}
\author[a]{Cathy Cheung}
\author[b]{Christine Ryan}
\author[b]{Darlene Lacey}
\author[b]{Elizabeth Kim}
\author[a]{Femi Olanubi}
\author[a]{Gabby Espinosa}
\author[b]{Gaby Andrews}
\author[a]{Gal Elidan}
\author[b]{Hannah Lieber}
\author[a]{Hairong Mu}
\author[b]{Jason Delgatto}
\author[b]{J.D. LaRock}
\author[b]{Jeannine Schloss}
\author[a]{Katherine Chou}
\author[a]{Nir Kerem}
\author[a]{Niv Efron}
\author[a]{Rajvi Kapadia}
\author[a]{Rena Levitt}
\author[a]{Rotem Yulzary}
\author[a]{Shashank Agarwal}
\author[a]{Tracey Lee-Joe}
\author[b]{Victor Martinez}
\author[a]{Yael Haramaty}
\author[a]{Yishay Mor}
\author[a]{Yossi Matias}
\newcommand{\sym}[1]{{#1}}

\affil[a]{Google Research}
\affil[b]{Network for Teaching Entrepreneurship}

\begin{abstract}
Durable skills like critical thinking, creativity, and collaboration are essential for innovation, yet teaching them remains difficult. In this work, we focus on the entrepreneurial domain, where these skills and others are required, and ask whether AI can be used as a tool for improving entrepreneurial proficiency. As a first step towards this challenge, we present an extension of Vantage, a platform developed at Google Research, adapted to serve as an interactive practice environment for entrepreneurship. Vantage pairs students with simulated AI peers engineered to navigate realistic business scenarios and challenge learners to exercise key skills. In a randomized controlled trial (RCT) embedded in summer entrepreneurship programs run by The Network for Teaching Entrepreneurship (NFTE), students using Vantage scored significantly higher on their Capstone Business Plans than control students with access to standard AI tools. These preliminary findings suggest that orchestrated AI peers can meaningfully support short-term performance on tasks requiring complex, durable skills.
\end{abstract}

\begin{document}
\maketitle

\section{Introduction}
The introduction of AI technology has emphasized the need to rethink the skills required in modern society and workplaces. Often referred to as 21st-century skills, or durable skills, they typically encompass
constructs such as creativity, critical thinking and collaboration, which play a key role in generating value and innovation in a rapidly evolving environment \cite{trilling200921st,voogt2012comparative,hilton2012education,oecd2019learningcompass}. 

Given the importance of durable skills, it is essential to develop a rigorous and effective pedagogical approach for their assessment and acquisition \cite{fiore2017collaborative,natalie2023innovating}. In recent work \cite{vantage_paper}, we introduced the Vantage approach for assessing these skills. Vantage is a platform that lets students interact with AI agents, and in the process elicits behavioral evidence of their durable skills.  The results in \cite{vantage_paper} suggested that the approach is effective at eliciting skill-related behavior, and that automatic rating of conversations is in good agreement with human rating of conversations. 
In this work, we address the complementary challenge of teaching durable skills. Our key insight is that an assessment system like Vantage can be augmented to function as a skill practice tool. This relies on the well-documented interplay between assessment and learning \cite{earl2012assessment,roediger2006test}.

One approach to skill teaching could be to focus on directly teaching isolated techniques for a specific durable skill. However, a key question is how these skills are actually deployed in practice. Addressing this requires a highly ecological learning environment in which the learner can practice these skills in scenarios that imitate real-world use. A natural setting for this kind of integrated application is entrepreneurship, which can be viewed as an overarching meta-skill that brings together various durable skills to pursue real-world goals \cite{wilson2009educating,obschonka2017entrepreneurship,ghafar2020convergence,peschl2021entrepreneurial,boyles201221st}. Notably, entrepreneurship education programs that require students to develop a culminating capstone project like a business plan contribute to the development of transferable durable skills. Within entrepreneurship education, business-plan development has traditionally been viewed as a key milestone in the venture creation process \cite{honig2004entrepreneurship,bewayo2015businessplans} and may foster higher-order skills such as problem-solving, communication, and self-reliance. Navigating the complex life cycle of business creation, from concept generation to iterative refinement, provides a realistic testing ground for skill application. 

{Indeed, the bi-directional link between entrepreneurship, education and durable skills is well acknowledged in educational policy and research \cite{ghafar2020convergence}: on the one hand, core skills such as critical thinking and creativity are essential for successful entrepreneurship. On the other hand, the context of entrepreneurship offers a useful organizing framework for these skills. It is important to note that entrepreneurship extends beyond business initiative, and includes social, scientific and cultural action and innovation \cite{jardim2021entrepreneurial}. As \cite{ghafar2020convergence} notes, traditional ``delivery'' mode methods in entrepreneurship education have been criticized as ineffective for skill development, and the prevailing approaches lean more towards the experiential, such as problem and project-based learning and simulations \cite{jardim2021entrepreneurial, rodrigues2023entrepreneurship}. In this context, there is a potential for using a simulation-based assessment tool such as Vantage as part of the learning experience: practicing skills in a safe environment and receiving detailed and timely feedback allows learners to explore and experiment with emerging skills without fear of negative consequences, and monitor their progress as they practice. Importantly, such simulations are not intended to replace authentic collaborative group work, but rather to augment it by offering advantages such as scalability and experimental control over learning scenarios.}

Given the goal of improving entrepreneurial skills, we modify Vantage to serve as an entrepreneurship practice environment, where students interact with AI peers to navigate realistic business scenarios, similar to tasks students encounter in entrepreneurship programs. The AI peers not only serve as high-realism teammates, but are also engineered to drive conversations into challenging scenarios that trigger skill utilization and generate practice and learning opportunities.

We evaluate the above system in a randomized controlled trial, carried out at an entrepreneurship summer program run by The Network for Teaching Entrepreneurship (NFTE). As the outcome variable we use the score on the final course project, where the students submit a business plan. Importantly, the project and its grading rubrics were developed by NFTE before and independently of our study.
The results show that students who used Vantage received higher scores on the business plan, as compared to students in the control group, who were given similar practice tasks and had access to standard generative AI tools, without the Vantage platform. These results suggest that an AI-based simulation environment such as Vantage can be used to improve performance on tasks that rely on core durable skills. 

\section{The Vantage System for AI-Based Practice of Durable Skills\label{sec:vantage_system}}
This section describes the Vantage system and its extension to practicing durable skills with AI peers.

\subsection{Vantage for Assessment}
Vantage is an experimental tool developed by Google Research \cite{vantage_paper} to allow students to demonstrate durable skills by interacting with AI teammates. The initial version of the system was focused on assessment of skills, and the AI teammates were driven by an Executive LLM designed to elicit information about a given set of focus skills (i.e., the skills that need to be assessed in the conversation). Results in \cite{vantage_paper} showed that the Executive LLM indeed elicited high rates of skill related behavioral evidence. In addition to providing a conversational interface, Vantage also uses AI to generate textual feedback on the skills and score the conversation according to rubrics. Data collected from human raters showed that the agreement between AI scores and human ratings was comparable to human inter-rater reliability.  

\subsection{Using Vantage to Practice Skills}

Although Vantage was initially designed with assessment in mind, it is natural to also use it for improving skills. First, its Executive LLM design drives the conversation with the student towards situations where skills need to be demonstrated, and thus prompts them to develop new strategies. Second, at the end of assessment sessions, actionable feedback informs students of areas for targeted improvement. Indeed, literature on formative assessment and test-enhanced learning demonstrates that structured assessment may foster skill acquisition \cite{earl2012assessment,roediger2006test}.

Motivated by the above, we modified the Vantage platform to facilitate skill practice. Specifically, we created an experience where users solve a set of tasks with AI peers. In what follows we provide details about the setup.

\begin{figure}[tbhp]
    \centering
    \includegraphics[width=0.85\linewidth]{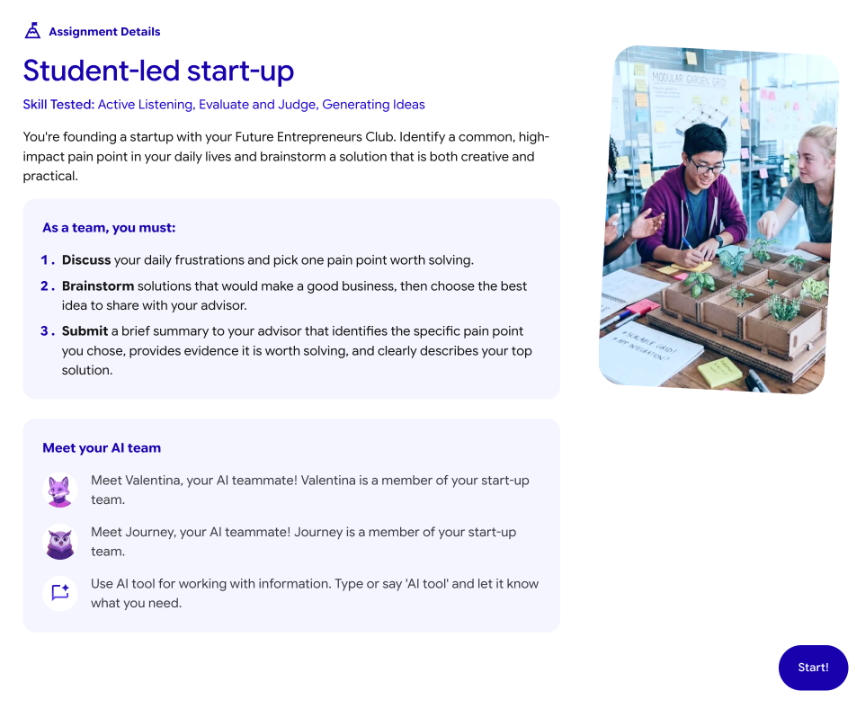}
    \vspace{-0.1in}
    \caption{The task description establishes the scenario, the student's role, the assignment details, and introduces the AI teammates.}
    \label{fig:vantage-task-brief}
\end{figure}

\begin{figure}[tbhp]
    \centering
    \vspace{-0.1in}
    \includegraphics[width=0.75\linewidth]{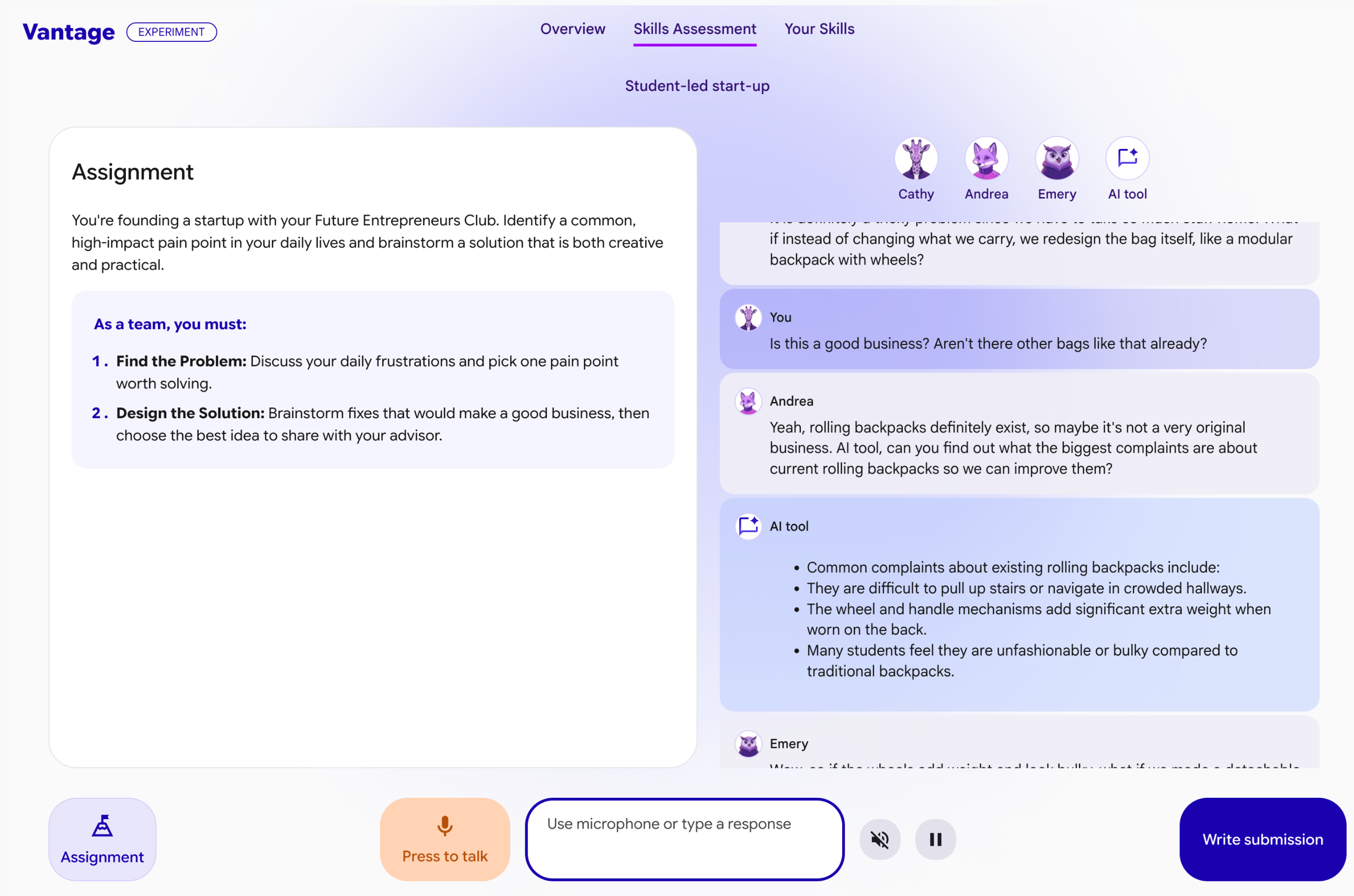}
    \caption{A session snippet of the student working through the task in conversation with two
    AI teammates and the in-platform AI tool.}
    \label{fig:vantage-task-convo}
\end{figure}

\subsubsection{The Vantage Skill Practice Session}

A Vantage session proceeds through the following four phases: setup, conversation, submission, and feedback. We next provide information about these, as well as about the Executive LLM that generates the text for the AI teammates.


\paragraph{Setup.} The student is placed in a real-life, authentic task and assigned a role within it (e.g., a financial consultant auditing a startup's expense list, or a marketing specialist helping founders articulate their brand purpose). The student is introduced to one or more AI teammates who occupy stakeholder roles in the same scenario: fellow founders, club members, members of a marketing team. The task brief (see \figref{fig:vantage-task-brief}) specifies what the team must accomplish and what the student will need to submit. See Table \ref{tab:six-tasks} in the sequel for the list of tasks used in the RCT. Each task is assigned a set of focus sub-skills that will be targeted in the conversation by the Executive LLM (see details below).

\paragraph{Conversation.} The student and the AI teammates work through the task in a conversation lasting roughly twenty minutes (see a snippet in \figref{fig:vantage-task-convo}). The teammates behave as collaborators rather than assistants: they contribute their own ideas, respond to the student's proposals, ask clarifying questions, and push back when they disagree. They do not complete the task on the student's behalf. An in-platform AI tool is available throughout for the student's independent use. The tool is instructed not to perform the cognitive work of the task or to supply direct answers; its role is to process information the student provides and to answer scenario-related factual questions.

\paragraph{Submission.} The session concludes when the student submits the assigned written deliverable. For example, this could be a customer profile, a pricing recommendation, or a corrected expense list.

\begin{figure}[tbhp]
    \centering
    \includegraphics[width=0.9\linewidth]{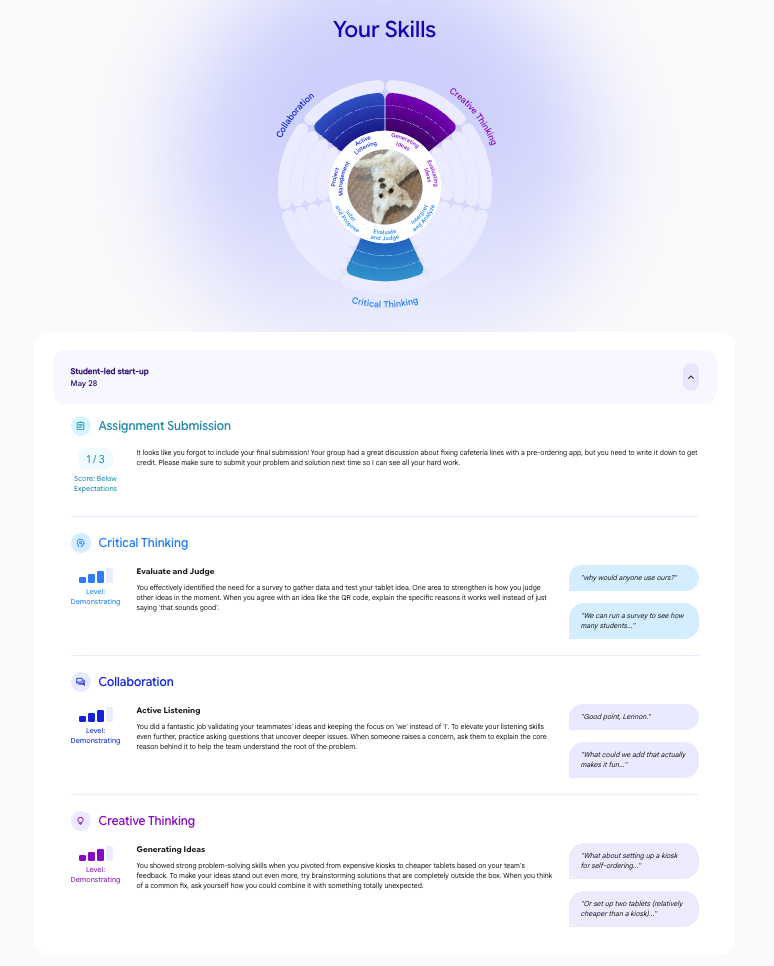}
    \caption{The feedback page returned after the student submits the task product. It shows the Vantage Submission Score; a Vantage Skill Score for each assessed sub-skill (Active Listening, Evaluate \& Judge, and Generating Ideas); and, for each sub-skill, a strength and a growth tip supported by quotes from the student's conversation.}
    \label{fig:feedback}
\end{figure}

\paragraph{Score and Feedback.} 
\ignore{
After submission, the student is presented with two types of score and feedback. The first is a Vantage Submission Score (1-3) for the written deliverable (e.g., a customer persona), together with a short feedback highlighting one strength and one recommendation. The second is Vantage Skill Scores (1–4) for all the assessed sub-skills based on the conversation, which always include the focus sub-skills, but also other sub-skills that are relevant for the task. In the experiments below the two focus sub-skills were always Evaluate \& Judge and Active Listening, and additional sub-skills were Generating Ideas, Infer \& Propose and Interpret \& Analyze.  Each Vantage Skill Score is accompanied by an strength and growth tips based on the conversation, and including quotes.\footnote{If the autorater concludes there is not enough information in the conversation about the skill, no feedback is provided.}
}

After submission, Vantage scores the session on two levels. The written deliverable receives a Vantage Submission Score, an integer from 1 to 3 assigned against the task's submission rubric, together with a short comment highlighting one strength and one recommendation. Each assessed sub-skill receives a Vantage Skill Score on a 1–4 rubric scale, based on evidence from the conversation; if the conversation does not contain enough evidence to score a sub-skill, it is marked N/A rather than assigned a level. Each Vantage Skill Score is accompanied by a strength and a growth tip, each supported by quotes from the student's own conversation (Figure \ref{fig:feedback}). Scores and feedback are shown to students on the six practice tasks but withheld on the pre- and post-tests. The scored skills always include the focus sub-skills, but also other sub-skills that are relevant for the task. In the experiments below the two focus sub-skills were always Evaluate \& Judge and Active Listening, and additional sub-skills were Generating Ideas, Infer \& Propose and Interpret \& Analyze.

\paragraph{The Executive LLM.}Turn generation for all AI teammates is orchestrated by a single Executive LLM that directs the conversation. The Executive LLM is the Gemini 3.1 Pro model prompted with general instructions, the task description, the rubrics for the sub-skills being assessed, and the conversation history. At each turn it chooses which teammate should speak — or none, so the student can lead — and generates that teammate's turn. The prompt instructs it to generate turns that drive the conversation toward obtaining information about the student's skill level, so that a reliable score can be inferred: it tracks the evidence gathered so far against each rubric level, identifies what is still missing, and plans in-character moves to elicit it. For example, for the Evaluate \& Judge skill, a teammate may suggest a flawed product idea that the student must evaluate, introduce a competing idea the student must reconcile, or state a false claim to test whether the student verifies it. 

At the same time, the prompt also instructs the AIs to ``behave naturally'' in the sense that their responses should be in character, and appear like a natural conversation, despite the model's behavior elicitation goal. They never adopt a teacher's or grader's tone, and never complete, solve, or submit the task for the student, even when the task is framed as a team deliverable. The Executive LLM also monitors the student's emotional state and eases off probing when the student shows signs of frustration or overload. The result is a conversation that serves two purposes: it mimics a realistic group interaction, and it exposes the student to a wide variety of challenges, leading to effective practice and skill acquisition.



\section{The Network for Teaching Entrepreneurship (NFTE) BizCamp}
In order to test the efficacy of the Vantage skill practice platform, Google Research partnered with The Network for Teaching Entrepreneurship (NFTE). NFTE is a global nonprofit organization founded in 1987 that provides entrepreneurship education to middle school, high school, and post-secondary learners. Its mission is to bring the power of entrepreneurship education to learners, educators, and decision-makers so that young people can develop the skills, knowledge, and experiences needed to own their futures. NFTE's educational model emphasizes entrepreneurial mindset development, experiential learning, opportunity recognition, problem solving, communication, and business creation through authentic, project-based experiences. NFTE delivers programming through schools, community-based organizations, and out-of-school learning environments across the United States and internationally. 

A central component of NFTE's educational approach is the development of entrepreneurial mindset, durable and transferable skills that support success in education, employment, and entrepreneurship. Towards this end, NFTE has developed many course offerings, ranging from two weeks to a full school year. The present study was conducted within NFTE BizCamps, an intensive entrepreneurship education program designed for middle and high school students. BizCamp provides a structured entrepreneurship learning experience in which students identify problems or opportunities, develop original business concepts, conduct market research, and create a business plan and/or venture pitch (Capstone Business Plan). It condenses the core learning of a year-long NFTEc course into a shorter period. Depending on local implementation, BizCamp programs range from approximately one to six weeks in duration and may be delivered through schools, community organizations, or other host institutions. NFTE also develops metrics for capturing aspects of entrepreneurship, including the Entrepreneurial Mindset Index (EMI), a validated self-report measure that captures student growth across key entrepreneurial mindset domains.

BizCamp combines entrepreneurship instruction with applied learning experiences intended to build entrepreneurial mindset, creativity, critical thinking, collaboration, decision-making, and communication skills. Students work through the process of developing a venture concept using NFTE's entrepreneurship curriculum and culminate the experience by presenting their business ideas to teachers, judges, or community stakeholders. The Capstone Business Plan produced during BizCamp served as the primary performance-based outcome measure for the present study.

\section{Experiment Design}
The Vantage platform (\secref{sec:vantage_system}) was deployed in the NFTE BizCamp within a randomized controlled trial, to test its efficacy as a skill practice platform. In this section we provide details about the experiment design. 

\subsection{Site Recruitment}

Recruitment for this study began in May of 2026, focusing on identifying BizCamp programs that were willing and able to implement the Vantage platform during the summer programming period and participate in study-related data collection activities. Nine BizCamp sites ultimately participated in the pilot study. Participating programs represented middle and high school settings located across the Southeast, Midwest, and West NFTE regions of the United States. Site participation was influenced by local implementation capacity, willingness to participate in research activities, technology access, and the ability to collect required outcome measures.

Student recruitment occurred within participating BizCamp programs. All enrolled students were eligible to participate in the research study, subject to parental consent and student assent procedures. 
Actual student participation fell below original projections because of variation in consent rates, implementation timelines, site-level logistics, and attendance patterns common to voluntary summer programming environments.

Because the study was implemented within existing BizCamp programs, students received entrepreneurship instruction regardless of study condition. Participation in the research study primarily influenced whether students received access to the Vantage intervention and whether their data contributed to specific study analyses.

\subsection{Randomized Controlled Trial Design}

The evaluation employed a mixed-methods design consisting of a randomized controlled trial (RCT) embedded within a broader feasibility and implementation study. The RCT was designed to estimate the effect of participation in Vantage. Additional treatment-only pilot sites were included to support implementation learning and exploratory analyses of platform performance.

Of the nine BizCamp sites, five implemented randomized assignment procedures. Four additional sites implemented Vantage as treatment-only pilots and were included in feasibility, implementation, and validity analyses but not causal impact analyses. Among the randomized sites, four ultimately contributed usable Capstone Business Plan outcome data for the primary impact analyses. 


Randomization occurred at the student level within participating classrooms rather than at the classroom or site level. This approach was selected to maximize statistical efficiency and reduce potential confounding associated with teacher, school, or site differences. Classrooms with at least 10 consented participants were eligible for random assignment. Within eligible classrooms, students were assigned to treatment or control conditions using an allocation ratio of approximately 65:35 favoring treatment. This allocation strategy was intentionally chosen to balance the goals of causal impact estimation with the need to generate sufficient usage data for implementation, product-development, and validation analyses.

Students assigned to the treatment condition were assigned six curriculum-aligned Vantage tasks, alongside the core BizCamp program. Students assigned to the control condition were assigned comparable entrepreneurship activities outside the Vantage platform while continuing to participate in the same core BizCamp program (see \secref{sec:task_design}).

Since Vantage can also serve to assess skills, an initial goal of the study was to check if the intervention improves Vantage assessment scores. Towards this end, students in all groups were expected to complete a Vantage assessment session, before and after the BizCamp. However, due to logistical and technical issues, across all sites, only 37 students completed both the pre- and post-Vantage test and only 24 of them received valid (non-N/A) Vantage Skill Scores on all assessed sub-skills in both tests. Thus, these assessment scores were not analyzed further.


\subsection{Task Design and Assignment \label{sec:task_design}}



Six tasks were developed jointly by Google Research and NFTE to use alongside the BizCamp curriculum. Each task was aligned to a specific BizCamp lesson so that the skills practiced in Vantage are exercised in the same conceptual territory the lesson covers. Both the control and treatment groups were assigned these tasks. In the treatment group, the task was described within the Vantage interface (See \figref{fig:vantage-task-brief} and \secref{sec:appendix_task_example}), and in the control group, the same task description was provided in a separate worksheet outside of Vantage. 

\begin{figure}

    \centering

    \includegraphics[width=0.9\linewidth]{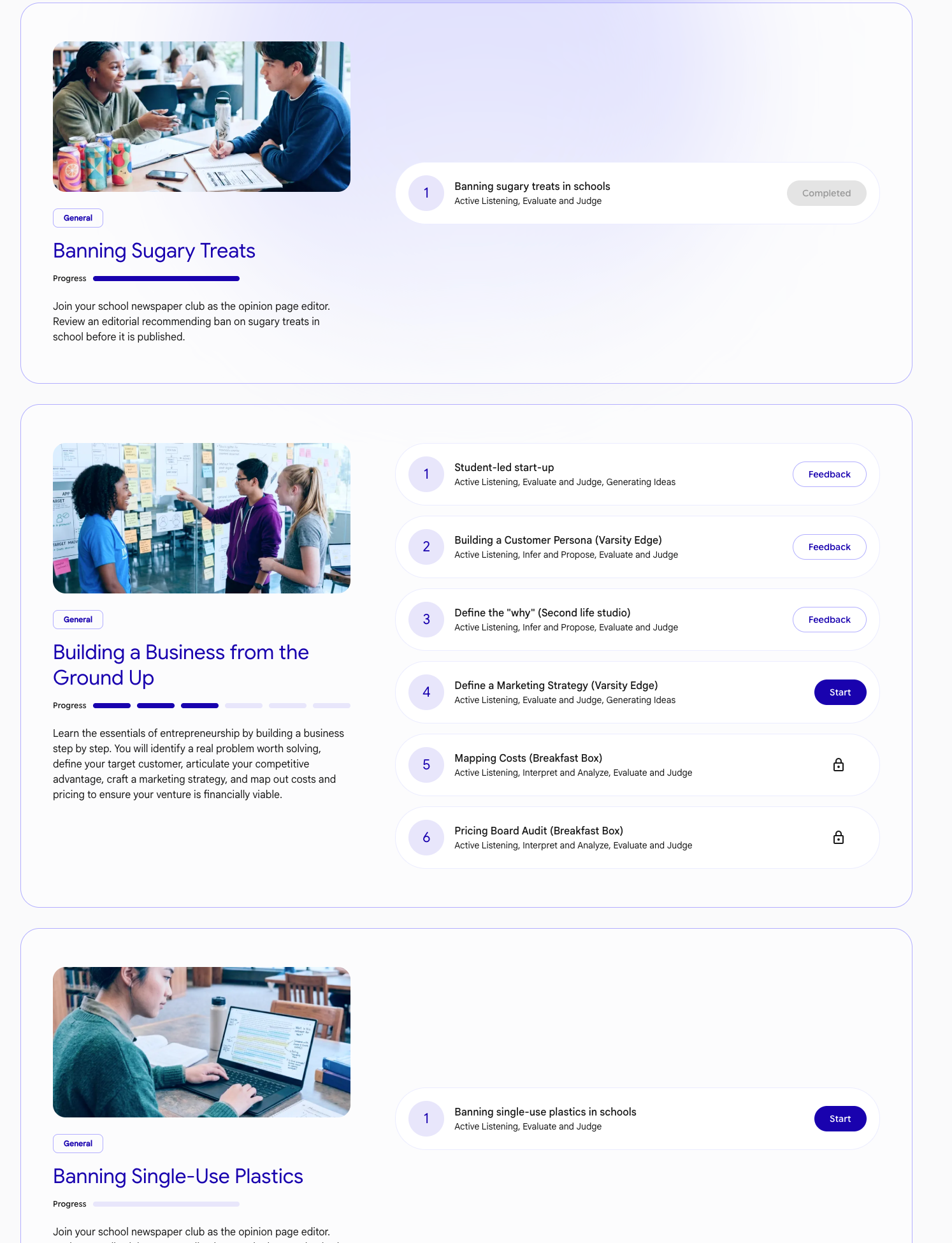}

    \caption{The Vantage home screen presented to treatment students. Three modules are shown: the pre-test (\textit{Banning Sugary Treats}), the six curriculum tasks grouped as \textit{Building a Business from the Ground Up}, and the post-test (\textit{Banning Single-Use Plastics}). Each task card lists the sub-skills on which it is assessing.}

    \label{fig:modules}

\end{figure}

Tasks were presented to students as a single module, \textit{Building a Business from the Ground Up}, in lesson order, and unlocked sequentially (see \figref{fig:modules}). Treatment students saw three modules on Vantage: pre-test (\textit{Banning Sugary Treats}), the six curriculum tasks grouped as \textit{Building a Business from the Ground Up}, and the post-test (\textit{Banning Single-Use Plastics}). Control students saw only the pre- and post-test modules on Vantage. See \secref{sec:appendix_pre_post_test}.

Every task was scored on three sub-skills. Two were common to all tasks: \textbf{Evaluate \& Judge} and \textbf{Active Listening}. The third was specific to the task, and was one of the three sub-skills: Generating Ideas, Infer \& Propose, or Interpret \& Analyze.  The assignment of these to the tasks was determined by the task content and is shown in Table \ref{tab:six-tasks}.

\begin{table}[htbp]

  \centering
\vspace{0.075in}
  \caption{The six curriculum-aligned Vantage tasks.}

  \label{tab:six-tasks}
\vspace{-0.075in}
  \scriptsize
  
  \begin{tabularx}{\linewidth}{@{}>{\raggedright\arraybackslash}p{2.3cm} >{\raggedright\arraybackslash}p{1.7cm} >{\raggedright\arraybackslash}p{5.0cm} >{\raggedright\arraybackslash}p{1.7cm} >{\raggedright\arraybackslash}p{4.2cm} X@{}}

    \toprule

    \textbf{NFTE Lesson} & \textbf{Vantage Task} &  \textbf{Scenario} &  \textbf{Task-Specific Sub-skill} &  \textbf{Deliverable} \\

    \midrule

    4: Business Ideas and Opportunities & Student-Led Startup & Brainstorm a business idea to solve a common pain point in students' daily lives  & Generating Ideas & Summary identifying a pain point, evidence it is worth solving, and a proposed solution \\

    \addlinespace

    7: Targeting a Customer Segment & Building a Customer Persona &  Help the founders of Varsity Edge — a student-run athletic coaching service — define their ideal customer & Infer \& Propose & Customer profile covering demographics, geographics, and buying patterns \\

    \addlinespace

    9: Competitive Advantage & Define the ``Why'' & Help the founders of Second Life Studio — an upcycled-apparel business — to articulate their core purpose & Infer \& Propose & Core purpose statement and three-frame storyboard outline \\

    \addlinespace

    11: Marketing Your Business & Define a Marketing Strategy & Work with Varsity Edge's marketing team to select marketing strategies  & Generating Ideas & One social media and two traditional marketing strategies with expected business impact \\

    \addlinespace

    13: The Cost of Doing Business & Mapping Costs & Audit the expense list drafted by the founders of Breakfast Box — a locker-delivery breakfast service & Interpret \& Analyze & Corrected expense list (startup / fixed / variable) with categorization rationale \\

    \addlinespace

    16: EOU and Financials & Pricing Board Audit & Review Breakfast Box's Economics-of-One-Unit analysis against customer survey data and recommend a selling price & Interpret \& Analyze & Pricing recommendation reconciling unit economics with target-market survey data \\


  \end{tabularx}

\end{table}


Students in the control condition were assigned six activities corresponding to the Vantage tasks in Table \ref{tab:six-tasks}. Each control activity presented the same scenario and required the same deliverable as its Vantage counterpart, but was completed individually using a worksheet rather than through conversation with AI teammates. Control students were permitted to use general-purpose AI tools of their choice while completing the activities.






\subsection{The Outcome Measure: Performance on Capstone Business Plan\label{sec:outcome_measures}}



The outcome measure used in the study was student performance on BizCamp Capstone Business Plans. Capstone Business Plans were evaluated using site-specific entrepreneurship rubrics by teachers and/or volunteer judges, and subsequently standardized as a percentage of possible points to support cross-site comparisons. Business plan scores were selected because they represented an authentic demonstration of the students' ability to apply entrepreneurship knowledge and related durable skills in a task that mirrors real-life entrepreneurial practice. 
  

\subsection{Self-Reported Learner \& Teacher Experience}

\label{sec:self_reported_experience}

To complement the quantitative analysis of Vantage's effect on capstone business plan performance, we conducted a follow-up mixed-methods evaluation of student and teacher experiences consisting of a post-intervention survey and semi-structured interviews.

\paragraph{Post-Intervention Student Survey}
In the final days of BizCamp, students across the RCT and treatment-only implementation study sites were invited to complete a user-experience survey (yielding $N = 51$ completed evaluations in Vantage and $N = 15$ in control). The instrument focused on two primary areas:
\begin{itemize}
    \item \textbf{Self-Reported Durable Skills Deployment:} Students' self-assessment of their own critical thinking, specifically their tendency to evaluate evidence, weigh alternative solutions, and assess peer arguments while using their assigned tools.
    \item \textbf{User Experience and Usability:} General perceptions regarding ease of use, task clarity, perceived utility toward their business projects, and overall enjoyment of the tools.
\end{itemize}


\section{Results}
Our key finding is that the Vantage intervention leads to a significant improvement in the outcome measure of business plan scores. We provide more details and analyses below.

\subsection{AI-Based Practice Improves Capstone Business Plan Scores}
This section reports the study's primary impact result: whether students who practiced with Vantage produced better Capstone Business Plans than students who did not. We first describe the analytic sample, then report the unadjusted treatment–control comparison and its consistency across sites, test whether the effect survives adjustment for site, and finally ask whether the effect scales with the amount of Vantage practice a student completed.

\subsubsection{Analytic Sample}
Of the nine participating sites, four (Sites 1, 2, 4, and 5) were treatment-only pilots with no control group and are excluded from impact analyses. Of the five randomized sites (Sites 3, 6, 7, 8, and 9), Site 8 has no Capstone Business Plan scores recorded and is also excluded. The four remaining randomized sites (Sites 3, 6, 7, and 9) constitute the impact sample. The resulting impact sample comprises 52 students: 36 treatment and 16 control.


\subsubsection{Treatment Advantage}
\begin{figure}
    \centering
    \vspace{-0.2in}
    \includegraphics[width=0.75\linewidth]{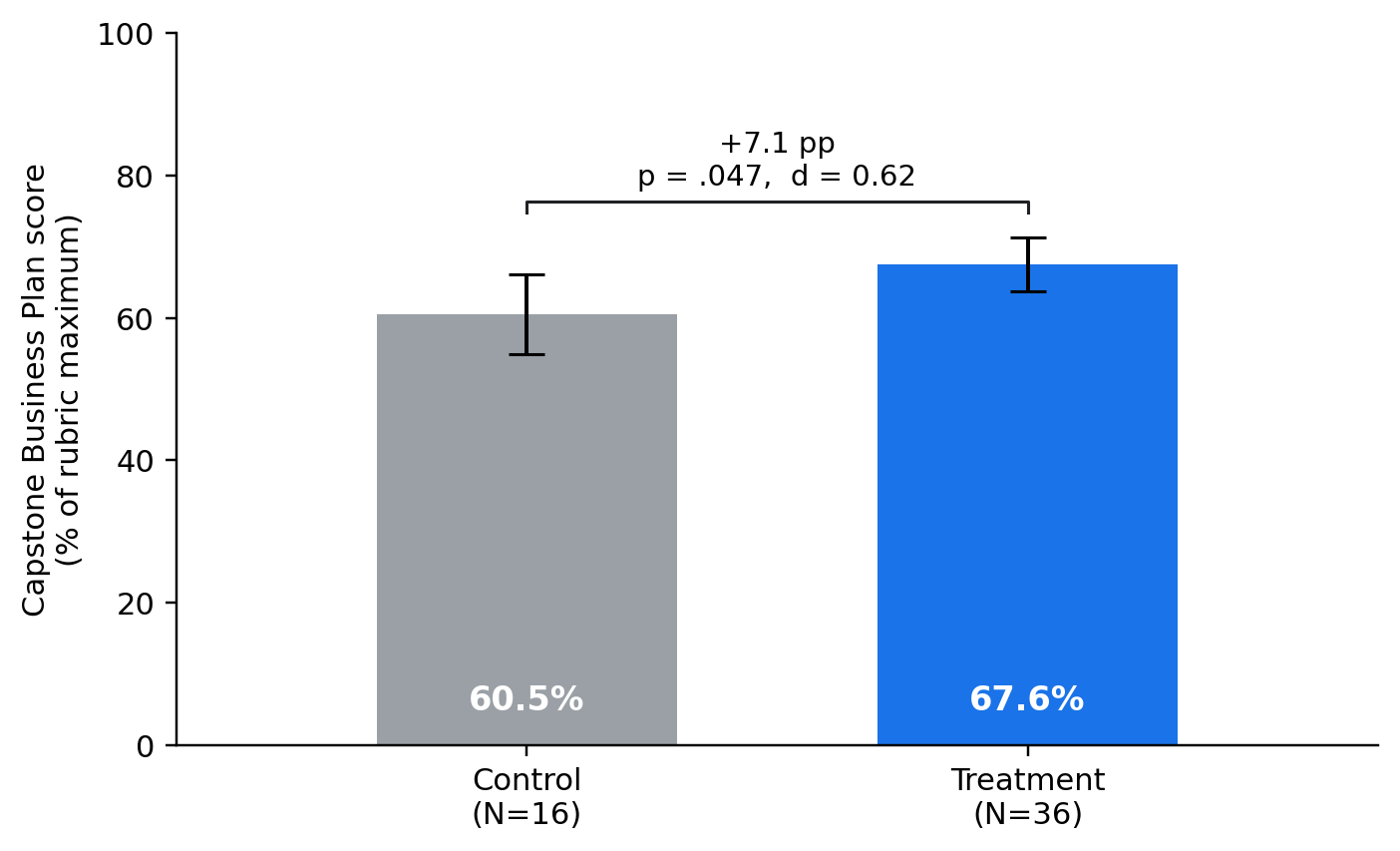}
    \caption{Mean Capstone Business Plan scores ($\pm 95\%\text{ CI}$) pooled across the four randomized sites ($N = 52$). Treatment students ($67.6\%, N = 36$) scored 7.1 percentage points higher than control students ($60.5\%, N = 16$; $p = .047$, Cohen's $d = 0.62$).}
    \label{fig:rct_bar_fig}
\end{figure}

\figref{fig:rct_bar_fig} shows the average Capstone Business Plan score for the treatment and control groups. Treatment students averaged $67.6\%$ (SD $11.5$, $N=36$) on the Capstone Business Plan against $60.5\%$ (SD $11.4$, $N=16$) for controls — a difference of $+7.1$ percentage points (Cohen's $d = 0.62$, $p = .047$). This effect exceeds the magnitude reported in prior entrepreneurship education research, where effect sizes on entrepreneurial mindset outcomes range from $d$ = $0.20$ to $0.32$ \cite{rodriguez2020mindset} and would be considered moderate-to-large according to conventional benchmarks \cite{cohen1988}. 

\begin{figure}[t]
    \centering
        \vspace{-0.2in}
    \includegraphics[width=0.8\linewidth]{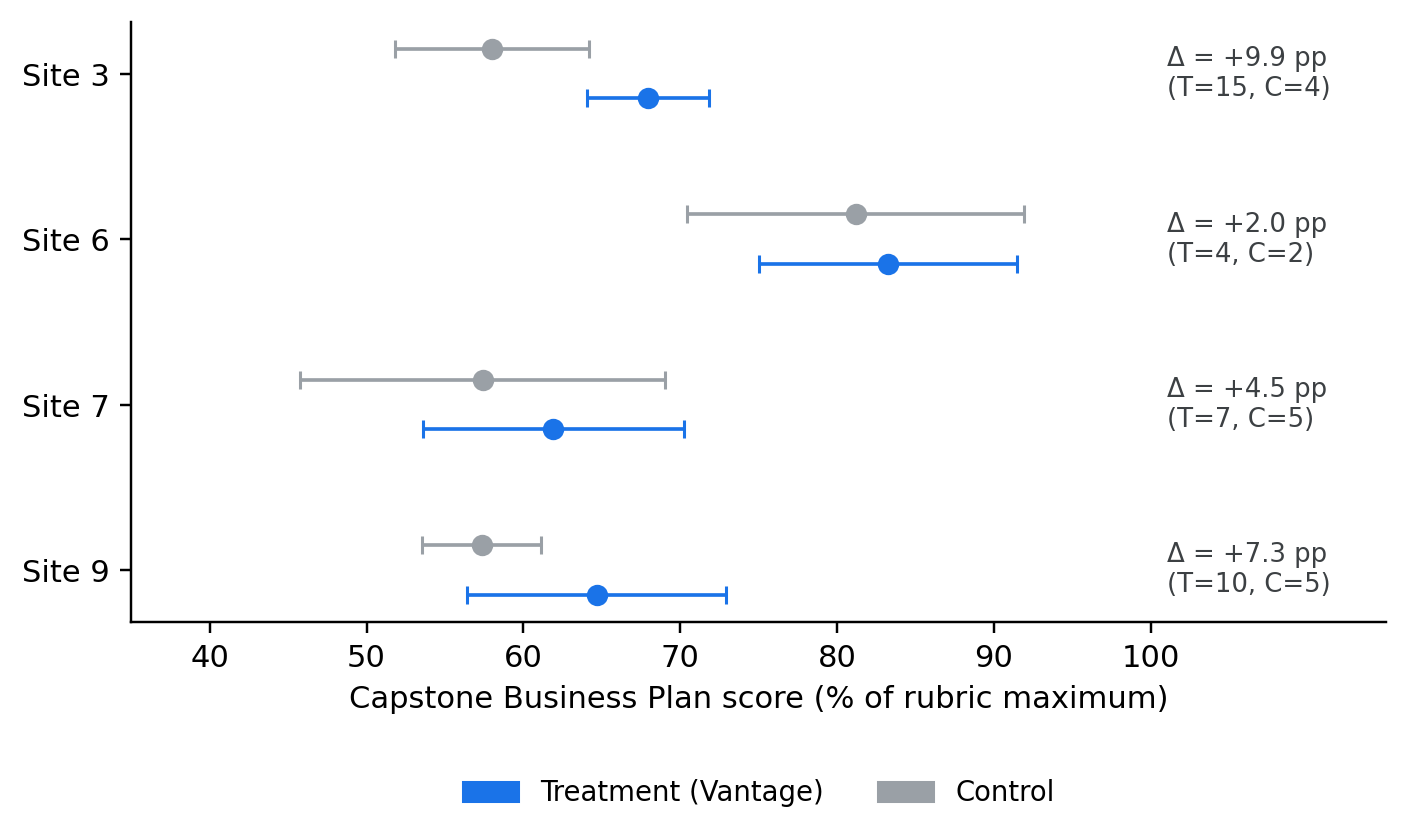}
    \caption{Mean Capstone Business Plan scores  within each randomized site. Treatment average scores exceeded control at every site, with differences ranging from +2.0 to +9.9 percentage points.}
    \vspace{0.05in}
    \label{fig:rct_sites_fig}
\end{figure}

The direction of the effect was consistent across all four randomized sites, with site-level differences ranging from $+2.0$ to $+9.9$ pp; no site showed control students outperforming treatment students. See \figref{fig:rct_sites_fig}. Per-site control groups contain only 2--5 students, however, so individual-site confidence intervals are wide and only Site 3 reaches significance on its own ($+9.9$ pp, $p = 0.030$).
Rather than serving as separate hypothesis tests, the site-level breakdown is presented to confirm that
the pooled effect is spread across sites rather than driven by one. Results reported here are for the pooled test, unless noted otherwise. 

\subsubsection{Effect Analysis under Covariate Adjustment}
To obtain a more precise estimate of the effect, we next adjust for the known covariate of site location. This is likely a source of outcome variance since factors such as course teachers and curriculum may affect business plan scores.
Table \ref{tab:ancova} shows the results, and the treatment effect remains positive and significant after adjustment. Adding site fixed effects raises R² from 8\% to 38\%, showing that site explains much of the variance in Capstone Business Plan scores. Since students were randomized within each classroom, site is unrelated to treatment assignment, the adjustment barely changes the treatment estimate, from +7.05 to +6.67 pp, while improving precision. 

\begin{table}
  \centering
  \caption{Treatment effect on Business Plan \% with and without site fixed effects (N = 52).}
  \label{tab:ancova}
  \small
  \begin{tabular}{@{}l r r r l r@{}}
    \toprule
    \textbf{Controls} & \textbf{$N$} & \textbf{$\beta$ (pp)} & \textbf{SE} & \textbf{$p$} & \textbf{$R^2$} \\
    \midrule
    None                                               & 52 & +7.05  & 3.46 & 0.047\sym{*}   & 7.7\%  \\
    Site fixed effects                                 & 52 & +6.67  & 2.96 & 0.029\sym{*}   & 38.2\% \\
    \bottomrule
  \end{tabular}
  \par\smallskip
  \begin{minipage}{0.1\linewidth}
    \footnotesize
    \sym{*}~$p<0.05$.
  \end{minipage}
\end{table}
%


\subsubsection{Dosage Analysis}
A natural follow-up question is whether treatment students who completed more curriculum-aligned Vantage tasks achieved higher Capstone Business Plan scores. To maintain consistency with the primary treatment advantage analysis, we examine dosage across the full randomized treatment cohort ($N=36$), counting students who completed no curriculum-aligned tasks as zero dosage. 

Across the six curriculum-aligned Vantage tasks, treatment students completed a mean of 2.61 tasks (median 3.0, range 0–6). Ten students completed 0 curriculum tasks and averaged 59.3\% on the Capstone Business Plan; the 5 students who completed 1–2 tasks averaged 65.7\%; the 16 students who completed 3–4 tasks averaged 69.5\%; and the 5 students who completed 5–6 tasks averaged 79.7\%. Being assigned to the treatment condition without practicing conferred no advantage relative to the control group (59.3\% vs. 60.5\%), whereas students who actively engaged with Vantage achieved progressively higher scores as they completed more of the curriculum.

Regressing Capstone Business Plan score on the number of completed curriculum-aligned tasks (0–6) across all 36 full randomized treatment students yields a statistically significant positive slope of +2.85 pp per task (SE $0.83$, $t = 3.43$, $p = .002$, R² = $25.8\%$). Controlling for site fixed effects, the dosage effect remains statistically significant at +2.33 pp per task (SE $0.87$, $t=2.68$, $p=.012$, R² = $41.2\%$), showing that the effect holds within classrooms rather than across sites. Thus, each additional curriculum-aligned Vantage task is associated with a 2.3 to 2.9 pp gain on the Capstone Business Plan.

\ignore{
\subsection{Vantage Submission Score Correlates with Capstone Business Plan Scores}
Within the treatment group, the average score students earned on their Vantage task submissions — the scores given to their written deliverable — was significantly associated with Capstone Business Plan performance ($r = +.38$, $p = .013$, $N=43$). Across all students with Vantage submission scores, including control students who completed the pre-/post-test, the association is stronger still ($r = +.48$, $p < .001$, $N=53$).

The Vantage Submission Score is the only Vantage measure that tracks Capstone Business Plan performance. Students' average Vantage Skill Scores across the practice tasks show no correlation with Capstone Business Plan scores (Active Listening $r = +.08$, $N=49$; Evaluate \& Judge $r = -.05$, $N=50$), and neither do their pre-test or post-test Vantage Skill Scores (all $|r| <= .22$, ns). 

Together with the dosage analysis in Section 5.1.4, this pattern indicates that how well students engage with the practice tasks matters more than how many tasks they complete: students who produced higher-quality written deliverables inside Vantage also produced stronger Capstone Business Plans at the end of camp.

\subsection{Grade-Level Effects}
The study sample spanned a wide developmental range, grades 6–12. We examined whether grade level is reflected in students' baseline measures and whether the Capstone Business Plan advantage reported in Section 5.1 holds for both younger and older students.

Because school grade and age were not recorded for every student, a \emph{derived grade} combining reported grade level and age was used. Students were grouped into middle school (grades 6–8) and high school (grades 9–12) bands. All analyses in this section exclude three engagement outliers (SID 35, adversarial sessions; SID 43, task-volume outlier; SID 102, idle-time outlier), leaving 87 students with a derived grade (23 middle school, 64 high school).

\subsubsection{School Grade Correlates With Vantage Pre-Test and Teacher Ratings}
Grade level is a statistically significant positive predictor of students' assessed skills at the start of camp. Derived grade correlates positively with Vantage Pre-test Skill Scores---Evaluate \& Judge ($r = +.43$, $p = .004$, $N = 44$) and Active Listening ($r = +.35$, $p = .028$, $N = 40$)---and with teacher ratings of problem solving, the teacher-rated counterpart\amirgc{What's this?} to Evaluate \& Judge ($r = +.35$, $p = .020$, $N = 43$). In contrast, school grade does not correlate with any Pre-program EMI or Post-program EMI scores (all $|r| < .15$, ns), consistent with students rating themselves relative to same-age peers. That the Vantage pre-test tracks grade level in the same direction as an independent teacher rating supports its validity as a measure of skill.\amirgc{Not sure we want to say that. By that argument grade level is also a good measure of skill...}

\subsubsection{Treatment Effect by Grade Band}
High school students have higher baseline Vantage Skill Scores: their mean pre-test Vantage Skill Score for Evaluate \& Judge was 2.43, compared with 1.50 for middle school students ($d = 0.09$, $p = .004$). Middle school treatment students engaged more with the Vantage platform, completing 5.53 Vantage tasks  on average compared with 3.91 Vantage task among high school treatment students ($d = -0.57$, $p = .045$, activity counts include the pre- and post-tests, the only activities available to control students).

Despite these differences, the treatment effect on the Capstone Business Plan was nearly identical across grade bands: +7.1 pp for middle school students (9 treatment, 3 control) and +6.8 pp for high school students (23 treatment, 12 control) at the four randomized sites, in line with the pooled +7.2 pp effect. Neither within-band estimate reaches significance on its own, and the middle school estimate rests on only three control students. The results provide no evidence that the benefit of practicing with Vantage is limited to younger or older students.

}

\subsection{Learner Experience}
\label{sec:results_learner_experience}
In what follows, we report results from the post-intervention survey.
Given the rate of completion for students who used Vantage across the RCT and implementation study sites ($N = 51$) and the control condition ($N = 15$), the analysis explores the self-reported experience of students who used Vantage across study sites. Results are provided in Table~\ref{tab:vantage_survey}.

\begin{table}
\centering
\footnotesize
\caption{Descriptive statistics from the post-intervention Vantage student survey ($N = 51$).}
\label{tab:vantage_survey}
\begin{tabular}{llcccc}
\toprule
\textbf{Dimension} & \textbf{Survey Prompt} & \textbf{\% Agree} & \textbf{\% Neutral} & \textbf{\% Disagree} \\
\midrule
\multirow{3}{*}{Critical Thinking} 
 & Evaluated info before decisions         & 70.5\% & 17.6\% & 11.8\% \\
 & Weighed different options               & 68.7\% & 19.6\% & 11.8\% \\
 & Assessed logic/arguments                & 70.0\% & 20.0\% & 10.0\% \\
\midrule
\multirow{2}{*}{Platform Dynamics} 
 & Feedback received was helpful           & 68.6\% & 21.6\% & 9.8\%\\
 & Partners were good teammates            & 54.9\% & 31.4\% & 13.7\%\\
\midrule
\multirow{3}{*}{Usability \& UX}   
 & Low difficulty & 72.5\%                       & ---    & 5.9\% \\
 & Overall utility     & 74.5\%    & ---    & 11.8\% (not at all)      \\
 & Overall enjoyment        & 58.9\%    & ---    & 25.5\% (not at all)      \\
\bottomrule
\end{tabular}
\footnotesize
\vspace{0.1in}
\end{table}

\noindent Three notable patterns emerged from the student survey responses:

\begin{enumerate}
    \item Students self-reported deep cognitive engagement with the material: a strong majority reported spending time evaluating evidence (70.5\%), weighing options (68.7\%), and assessing arguments (70.0\%) during tasks.
    \item Most participants found the automated feedback from AI teammates helpful (68.6\%), but they were more divided on characterizing their teammates as ``good'' (54.9\%)---affirming the product design choice for AI teammates to be intentionally semi-helpful rather than passive assistants.
    \item 
    Student feedback indicated high usability (72.5\% rated it low difficulty) and high perceived utility toward BizCamp (74.5\%). However, participants were divided on their overall enjoyment: 58.9\% reported positive enjoyment, while 25.5\% stated they did not enjoy the tool at all. This again can be partially attributed to the fact that Vantage AI peers present more friction than AI assistants, which may affect the overall perception of the experience.
\end{enumerate}

Taken together, these results suggest a promising pattern: Vantage was easy to use and was associated with high levels of self-reported cognitive engagement during collaboration with AI peers. However, the experience was not universally enjoyable, and students did not always view their AI partners as consistently capable or agreeable. While further investigation is needed, these findings highlight an opportunity to better balance productive friction with learner satisfaction—specifically by examining how intentional challenge can stimulate critical thinking without leading to frustration or diminished engagement.



\section{Discussion}
Artificial intelligence has the potential to improve and augment existing teaching practices, and open new opportunities for teaching and learning. The natural first steps towards this goal were for assessment and tutoring of ``classic'' school curricula such as math and sciences \cite{team2026teaching}. These have been shown to be effective in some cases \cite{team2026teaching,kestin2025ai}, although other studies provide mixed results (see \cite{bastani2025generative} and a review in \cite{fesler2026evidence}). 

Here, we shift the focus to the challenge of teaching durable skills at scale. While these skills are important for standard curriculum learning, they are equally necessary across a wide range of broader activities, from collaborative group work to content creation. It is thus a timely goal to introduce effective approaches to measurable  improvement of these skills.
In this setting, there are relatively few standardized assessments\footnote{Of the 21st century skills, the one that has been most amenable to standardization is probably critical thinking, where assessments such as Watson Glaser\texttrademark\ or the Cornell Critical Thinking Test (see \cite{ennis1993critical} for a review) have been developed.}, and while teaching methods have been developed, they have been challenging to study and apply at scale (e.g., see \cite{abrami2015strategies} for teaching critical thinking). 

In this work, we argue that an environment where students are required to elicit skill-related behavior can also naturally act as a practice environment for skill-building. 
While our Vantage environment does not explicitly introduce a skill teaching curriculum, it lets students practice these skills across six different tasks with AI agents that are instructed to behave naturally while challenging students to apply several skills. Thus, the AI agents act as teammates rather than assistants or tutors. 
The results of the RCT suggest that this approach can have a positive and statistically significant effect on the course outcome. 

While this is a promising first step, much work remains to be done to develop AI skill-tutoring systems and demonstrate their efficacy. As noted above, our approach does not rely on a teaching curriculum and does not explicitly target the many facets of durable skills (e.g., understanding causality in critical thinking). Furthermore, our current intervention lasted a few weeks, whereas a longer period may prove more effective. Finally, future work could explore additional aspects such as cultural adaptation, motivation, and alternating assessment and practice. 

The RCT reported here shows improvement on an important outcome variable: the scores of a capstone project. While the project is not a direct measurement of entrepreneurship or specific durable skills, it is likely to be correlated with these. Thus, showing that it improves suggests that Vantage practice has a positive effect on these skills. 
However, empirically establishing the broader efficacy of these intervention methods requires larger and more diverse samples, and the measurement of more outcome variables to quantify improvement on different skills. 


The results in this study, while preliminary, suggest that AI-supported practice may be a promising approach for developing durable skills in students. 
By lowering the barriers to personalized, experiential learning, this approach offers a promising model for helping students build the competencies required for an evolving society.
\bibliography{skills}

\newpage
\section{Appendix}
This appendix provides additional information about the tasks used in the RCT.
\subsection{Vantage Curriculum-aligned Tasks}
\label{sec:appendix_task_example}
Table \ref{tab:nfte_tasks} provides additional information about the task description and submission instructions for the six tasks that students in the NFTE RCT received. 

{\small
\setlength{\tabcolsep}{6pt}
\renewcommand{\arraystretch}{1.25}
\begin{longtable}{@{} >{\raggedright\arraybackslash\bfseries}p{0.18\textwidth} >{\raggedright\arraybackslash}p{0.48\textwidth} >{\raggedright\arraybackslash}p{0.28\textwidth} @{}}
\caption{Overview of NFTE Tasks, Descriptions, and Submission Instructions.}
\label{tab:nfte_tasks} \\
\toprule
\textbf{Task Title} & \textbf{Task Description} & \textbf{Submission Instruction} \\
\midrule
\endfirsthead
\multicolumn{3}{c}{{\tablename\ \thetable{} -- continued from previous page}} \\
\toprule
\textbf{Task Title} & \textbf{Task Description} & \textbf{Submission Instruction} \\
\midrule
\endhead
\midrule
\multicolumn{3}{r}{{Continued on next page}} \\
\bottomrule
\endfoot
\bottomrule
\endlastfoot
Student-led start-up &
You're founding a startup with your Future Entrepreneurs Club. Identify a common, high-impact pain point in your daily lives and brainstorm a solution that is both creative and practical.\par\medskip
As a team, you must:
\begin{enumerate}[nosep, leftmargin=1.5em, topsep=2pt]
    \item \textbf{Discuss} your daily frustrations and pick one pain point worth solving.
    \item \textbf{Brainstorm} solutions that would make a good business, then choose the best idea to share with your advisor.
    \item \textbf{Submit} a brief summary to your advisor that identifies the specific pain point you chose, provides evidence it is worth solving, and clearly describes your top idea.
\end{enumerate} &
Write a 3 to 5 sentence paragraph that:
\begin{enumerate}[nosep, leftmargin=1.5em, topsep=2pt]
    \item Identifies the specific pain point you chose.
    \item Provides evidence that it is worth solving.
    \item Describes your top idea to directly address the pain point.
\end{enumerate} \\
\midrule
Building a Customer Persona (Varsity Edge) &
You are a business development specialist. Varsity Edge is a student-run service where high school athletes give 1-on-1 training to middle school students who want to improve their skills.\par\smallskip
Explore the goals and frustrations of athletes and parents to see who will pay for special coaching. Work with the founders to describe their ideal customer so the persona feels realistic.\par\medskip
As a team, you must:
\begin{enumerate}[nosep, leftmargin=1.5em, topsep=2pt]
    \item \textbf{Define} demographics (age, education, occupation, income) and geographics (city, state, neighborhood) for your ideal customer.
    \item \textbf{Describe} their psychographics (motivations and goals, frustrations) and buying patterns (method and frequency).
    \item \textbf{Submit} a customer profile for the persona you built. Make sure to include their demographics, geographics, psychographics, and buying patterns.
\end{enumerate} &
Outline a detailed customer persona profile that includes:
\begin{enumerate}[nosep, leftmargin=1.5em, topsep=2pt]
    \item All necessary demographics (age, education, occupation, income).
    \item Specific geographics (city, state, neighborhood).
    \item Psychographics details describing their personality and preferences (motivations, goals, frustrations).
    \item Buying patterns (method and frequency).
\end{enumerate} \\
\midrule
Define the ``why'' (Second life studio) &
You are a marketing expert who works with eco-friendly businesses. You are working with Second Life Studio. This new business turns fabric waste into premium apparel. They partner with regional thrift centers to collect unwearable clothing. The clothing is recycled into durable fabric for a line of premium, zero-waste outerwear. Help them define their core purpose and communicate it to customers.\par\medskip
As a team, you must:
\begin{enumerate}[nosep, leftmargin=1.5em, topsep=2pt]
    \item \textbf{Explore} impact models like local job creation or conscious sourcing for Second Life Studio. Select one that defines the studio's core purpose beyond profit. This is their ``why''.
    \item \textbf{Outline} a 3-part storyboard that communicates Second Life Studio's core purpose to their target customers.
    \item \textbf{Submit} a core purpose statement explaining ``why'' Second Life Studio exists and your storyboard outline for communicating it with customers.
\end{enumerate} &
Write a brief explanation of Second Life Studio's ``why'' that includes:
\begin{enumerate}[nosep, leftmargin=1.5em, topsep=2pt]
    \item A 1--2 sentence core purpose statement.
    \item An outline of your 3-part storyboard that communicates this core purpose to customers.
\end{enumerate} \\
\midrule
Define a Marketing Strategy (Varsity Edge) &
You are a marketing consultant working with Varsity Edge. This student-run coaching service features high school athletes who provide 1-on-1 personal training to middle school students.\par\smallskip
Work with their marketing team to develop three plans. You need two traditional marketing strategies and one social media strategy that will reach middle school athletes and their families.\par\medskip
As a team, you must:
\begin{enumerate}[nosep, leftmargin=1.5em, topsep=2pt]
    \item \textbf{Discuss} different marketing approaches that will reach students and their families.
    \item \textbf{Select} a social media strategy and 2 traditional methods. For each tactic, explain its purpose, how it will reach customers and its business impact.
    \item \textbf{Submit} a description of the 3 marketing strategies your team selected for Varsity Edge. Describe how each strategy will reach customers and impact the business.
\end{enumerate} &
Describe marketing strategies you selected for Varsity Edge. Include:
\begin{enumerate}[nosep, leftmargin=1.5em, topsep=2pt]
    \item A description of your social media strategy that is specific to the business.
    \item A description of your 2 traditional strategies, how they will reach customers, and how they will impact the business.
\end{enumerate} \\
\midrule
Mapping Costs (Breakfast Box) &
You are a financial strategy consultant. You are working with the student founders of Breakfast Box. This business delivers nutritious, grab-and-go meals directly to student lockers.\par\smallskip
Their team drafted a list of startup, fixed, and variable expenses. Work with their team to audit their draft before they finalize their business model.\par\medskip
As a team, you must:
\begin{enumerate}[nosep, leftmargin=1.5em, topsep=2pt]
    \item \textbf{Review} the founders' lists of expenses. Identify items that have been miscategorized, like a startup cost labeled as a variable cost.
    \item \textbf{Identify} and categorize any missing expenses from the list.
    \item \textbf{Submit} an updated list with your corrections and additions. Include an explanation of how to categorize expenses to help the Breakfast Box team correctly generate future reports.
\end{enumerate} &
Provide your updates for Breakfast Box. Include:
\begin{enumerate}[nosep, leftmargin=1.5em, topsep=2pt]
    \item Your corrected expense list.
    \item An explanation of how to categorize expenses for future reports.
\end{enumerate} \\
\midrule
Pricing Board Audit (Breakfast Box) &
You are a financial consultant for a student business incubator. You are working with the founders of Breakfast Box. This business delivers nutritious, grab-and-go meals directly to student lockers.\par\smallskip
They submitted an Economics of One Unit (EOU) analysis. Your job is to review the analysis, suggest updates to make sure the selling price is fair and will make enough profit.\par\medskip
You must:
\begin{enumerate}[nosep, leftmargin=1.5em, topsep=2pt]
    \item \textbf{Check} the startup's financial calculations and determine whether their cost estimates are realistic.
    \item \textbf{Evaluate} unit price and profit. Should the price go up, down, or stay the same? Do they need to reconsider some expenses? This could include finding cheaper supplies or changing the delivery fees.
    \item \textbf{Submit} a summary of your recommended changes. Explain the reasoning for your recommendations.
\end{enumerate} &
Summarize your recommendations for Breakfast Box. Include:
\begin{enumerate}[nosep, leftmargin=1.5em, topsep=2pt]
    \item Any issues you identified with the estimated costs and how they should be fixed.
    \item The price you recommend and why, balancing market fairness with profit.
\end{enumerate} \\
\end{longtable}
}

\newpage
\subsection{Vantage Pre- and Post- Test}
\label{sec:appendix_pre_post_test}
In Table \ref{tab:nfte_pre_post_tasks} we provide details about the Vantage assessment tasks (pre and post) that both control and treatment groups took in the RCT.

{\small
\setlength{\tabcolsep}{6pt}
\renewcommand{\arraystretch}{1.25}
\begin{longtable}{@{} >{\raggedright\arraybackslash\bfseries}p{0.18\textwidth} >{\raggedright\arraybackslash}p{0.48\textwidth} >{\raggedright\arraybackslash}p{0.28\textwidth} @{}}
\caption{Overview of NFTE Pre-Test and Post-Test Assessment Tasks.}
\label{tab:nfte_pre_post_tasks} \\
\toprule
\textbf{Task Title} & \textbf{Task Description} & \textbf{Submission Instruction} \\
\midrule
\endfirsthead
\multicolumn{3}{c}{{\tablename\ \thetable{} -- continued from previous page}} \\
\toprule
\textbf{Task Title} & \textbf{Task Description} & \textbf{Submission Instruction} \\
\midrule
\endhead
\midrule
\multicolumn{3}{r}{{Continued on next page}} \\
\bottomrule
\endfoot
\bottomrule
\endlastfoot
Banning sugary treats in schools (Pre-test) &
You are the editor of your school newspaper's opinion section. Two AI student journalists authored an article draft recommending banning the sale of candy and soda on campus. You must evaluate the article's reasoning and evidence to decide whether it should be published.\par\medskip
As the editor, you must:
\begin{enumerate}[nosep, leftmargin=1.5em, topsep=2pt]
    \item \textbf{Review} the article. Look for any questionable claims or flawed reasoning.
    \item \textbf{Meet} with the AI student journalists who authored the article to discuss their evidence and reasoning. Use your AI tool to help fact-check in real time.
    \item \textbf{Submit} a short message to your head editor with your decision whether to publish or reject the current article, a justification of your decision, and feedback to the AI student journalist authors.
\end{enumerate} &
Write a short message to your head editor with your decision whether to publish or reject the current article, a justification of your decision, and feedback to the AI student journalists who authored it. \\
\midrule
Banning single-use plastics in schools (Post-test) &
You are the editor of your school newspaper's opinion section. Two AI student journalists authored an article draft recommending banning single-use plastics in the school cafeteria. You must evaluate the article's reasoning and evidence to decide whether it should be published.\par\medskip
As the editor, you must:
\begin{enumerate}[nosep, leftmargin=1.5em, topsep=2pt]
    \item \textbf{Review} the article. Look for any questionable claims or flawed reasoning.
    \item \textbf{Meet} with the AI student journalists who authored the article draft to discuss their evidence and reasoning. Use your AI tool to help fact-check in real time.
    \item \textbf{Submit} a short message to your head editor with your decision whether to publish or reject the current article, a justification of your decision, and feedback to the AI student journalist authors.
\end{enumerate} &
Write a short message to your head editor with your decision whether to publish or reject the current article, a justification of your decision, and feedback to the AI student journalists who authored it. \\
\end{longtable}
}

\end{document}